\documentclass[pdftex,twocolumn,epjc3]{svjour3}  

\usepackage{lineno}
\usepackage[utf8]{inputenc}
\usepackage[english]{babel}
\usepackage[euler]{textgreek}
\usepackage{amsmath}
\usepackage{color}
\usepackage{amssymb}
\usepackage{mathtools,cuted}
\usepackage{feynmp}
\usepackage{feynmp-auto}
\usepackage{slashed}
\usepackage{float}
\usepackage{physics}
\usepackage{tikz}
\usepackage[export]{adjustbox}

\RequirePackage[T1]{fontenc}

\smartqed  

\RequirePackage{graphicx}
\RequirePackage{mathptmx}      
\RequirePackage{flushend}
\RequirePackage[numbers,sort&compress]{natbib}
\RequirePackage[colorlinks,citecolor=blue,urlcolor=blue,linkcolor=blue,draft]{hyperref}

\journalname{Eur. Phys. J. C}

\begin{document}

\title{Operation of a ferromagnetic Axion haloscope at $m_a=58\,\mu\mathrm{eV}$}

\author{N. Crescini\thanksref{e1,addr1,addr2}
        \and
        D. Alesini\thanksref{addr3}
        \and
        C. Braggio\thanksref{addr1,addr4}
        \and
        G. Carugno\thanksref{addr1,addr4}
        \and
        D. Di Gioacchino\thanksref{addr3}
        \and
        C. S. Gallo\thanksref{addr2}
        \and
        U. Gambardella\thanksref{addr6}
        \and
        C. Gatti\thanksref{addr3}
        \and
        G. Iannone\thanksref{addr6}
        \and
        G. Lamanna\thanksref{addr7}
        \and
        C. Ligi\thanksref{addr3}
        \and
        A. Lombardi\thanksref{addr2}
        \and
        A. Ortolan\thanksref{addr2}
        \and
        S. Pagano\thanksref{addr6}
        \and
        R. Pengo\thanksref{addr2}
        \and
        G. Ruoso\thanksref{e2,addr2}        
        \and
        C. C. Speake\thanksref{addr8}
        \and
        L. Taffarello\thanksref{addr4}        
}

\thankstext{e1}{e-mail: nicolo.crescini@phd.unipd.it}
\thankstext{e2}{e-mail: ruoso@lnl.infn.it}

\institute{Dipartimento di Fisica e Astronomia, Via Marzolo 8, I-35131 Padova (Italy)\label{addr1}
          \and
          INFN, Laboratori Nazionali di Legnaro, Viale dell'Universit\`a 2, I-35020 Legnaro, Padova (Italy)\label{addr2}
          \and
		  INFN, Laboratori Nazionali di Frascati, Via Enrico Fermi, 40, I-00044 Frascati, Roma (Italy)\label{addr3}
		  \and
		  INFN, Sezione di Padova, Via Marzolo 8, I-35131 Padova (Italy)\label{addr4}
		  \and
		  INFN, Sezione di Napoli and University of Salerno, Via Giovanni Paolo II 132, I-84084 Fisciano (Italy)\label{addr6}
		  \and
		  INFN Sezione di Pisa and University of Pisa, Largo Bruno Pontecorvo 3, 56127 Pisa (Italy)\label{addr7}
		  \and
		  School of Physics and Astronomy, University of Birmingham, West Midlands B15 2TT (UK)\label{addr8}
}

\date{Received: date / Accepted: date}

\maketitle

\begin{abstract}
Axions, originally proposed to solve the strong CP problem of quantum chromodynamics, emerge now as leading candidates of WISP dark matter. The rich phenomenology associated to the light and stable QCD axion can be described as an effective magnetic field that can be experimentally investigated. 
For the QUAX experiment, dark matter axions are searched by means of their resonant interactions with electronic spins in a magnetized sample. In principle, axion-induced magnetization changes can be detected by embedding a sample in an rf cavity in a static magnetic field. In this work we describe the operation of a prototype ferromagnetic haloscope, with a sensitivity limited by thermal fluctuations and receiver noise. 
With a preliminary dark matter search, we are able to set an upper limit on the coupling constant of DFSZ axions to electrons $g_{aee}<4.9\times10^{-10}$ at 95\% C.L. for a mass of $58\,\mu$eV (i.\,e. 14\,GHz). This is the first experimental result with an apparatus exploiting the coupling between cosmological axions and electrons.
\end{abstract}

\section{Introduction}
\label{intro}
A major fraction of the mass content of the universe is composed of dark matter (DM), i.e. particles not interacting significantly with electromagnetic radiation, with ordinary matter or self-interacting (cold dark matter) \cite{zwicky,rubin78,rubin80}.
Up-to-date results \cite{pdg} show that with respect to the universe critical density the DM fraction is the $25.8\%$ while the luminous matter fraction is $5.7\%$, meaning that DM is about five times more abundant than ordinary baryonic matter.
This outstanding result triggered theoretical studies aiming to understand the nature of DM, for instance in the form of new particles beyond the Standard Model (SM). 

The axion is a good candidate for DM but was not originally introduced to account for this specific issue. To solve the strong CP problem Peccei and Quinn added a new symmetry to the SM \cite{pq}, which breaks at an extremely high energy scale $F_a$ producing a pseudo-Goldstone boson, the axion \cite{weinberg1978new}.
Among the proposed models, the ``invisible axion'' model classes KSVZ and DFSZ still hold \cite{DINE1981199,SHIFMAN1980493,PhysRevLett.43.103,DINE1983137}.
For scales $F_a\sim10^{12}\,$GeV, corresponding to typical mass values  $m_a\lesssim1\,$meV, large quantities of axions may have been produced in the early universe and could account even for the totality of cold dark matter \cite{wilczek1978problem}. 
Consequently, several detection schemes have been devised during the last decades to search for relic axions. 
The value of $F_a$ is not fixed by the theory, however, cosmological considerations and astrophysical observations \cite{PRESKILL1983127,raffelt1996stars,turner1990windows,abbott1983cosmological,fox2004probing,spergel170wilkinson} provide boundaries on $F_a$ and suggest a favoured axion mass range $1\,\mu\mathrm{eV}<m_a<10\,\mathrm{meV}$.
In addition, lattice results on QCD topological susceptibility, based on reliable computations of the axion relic density, indicate a preferred window for the axion mass in the range of tens of $\mu$eV 
\cite{BURGER2017880,berkowitz2015lattice,borsanyi2016calculation,diCortona2016,petreczky2016topological,bonati}.

Axion model classes can be tested with different experimental techniques \cite{PhysRevLett.51.1415,ringwald,axion_searches,redondo,kim}.	
Most of these experiments are based on the Primakoff effect, i.e. an axion to photon conversion in a strong static magnetic field \cite{PhysRevLett.112.091302,1475-7516-2007-04-010,PhysRevLett.118.061302,PhysRevLett.104.041301,PhysRevD.74.012006,MCALLISTER201767,refId0,caldwell2017dielectric}. In particular, the ADMX experiment reached the cosmologically relevant sensitivity to exclude the axion mass range $1.9\,\mu\mathrm{eV}\lesssim m_a\lesssim3.7\,\mu\mathrm{eV}$ for the KSVZ model and $2.66\,\mu\mathrm{eV}<m_a<2.81\,\mu\mathrm{eV}$ for the DFSZ model \cite{PhysRevLett.120.151301}, assuming a local DM density of 0.45\,GeV/cm$^3$.
On the other hand the axion-fermion coupling, explicitly predicted in different axion models including DFSZ \cite{PhysRevLett.118.071802,1475-7516-2017-08-001,Ernst2018,Ema2017,PhysRevD.95.095009}, allows for designing new detectors that exploit the interaction between axions and fermionic spins \cite{Garcon:2017ixh,PhysRevX.4.021030,1742-6596-718-4-042051}. Among these, the QUAX detector \cite{BARBIERI2017135} takes advantage of the resonant interaction between relic axions and a magnetized magnetic sample housed in a microwave cavity.
In this paper we present results on the operation of a QUAX demonstrator, based on 5 GaYIG (Gallium Yttrium Iron Garnet) 1\,mm diameter spheres placed in a 14\,GHz resonant cavity. The apparatus is operated at cryogenic temperatures and its sensitivity is limited only by thermal effects.
Section \ref{esr} describes the proposed detection scheme, Sections \ref{quax_prototype} and \ref{discussioni} report on the measurement of an upper limit on the axion interaction with electronic spins, using the small-scale prototype of the final apparatus. Conclusions are eventually drawn in Section \ref{conclusion}.

\section{Axion detection by resonant interaction with electron spin}
\label{esr}
The first ideas on axion detection via their conversion to magnons, collective excitations of the spins in a ferromagnet, were discussed in Ref.s~\cite{vorobyov1995ferromagnetic,BARBIERI1989357,kakhidze1991antiferromagnetic,Caspers:1989ix}.
As the DFSZ axion and other axion models \cite{PhysRevLett.118.071802,1475-7516-2017-08-001,Ernst2018,Ema2017,PhysRevD.95.095009} does not suppress the coupling between an axion $a$ and an electron $\psi$ at the tree level, the Lagrangian reads
	\begin{equation}
	{\cal L}=\bar{\psi}(x)(i\hbar 
\gamma^\mu \partial_\mu- mc)\psi(x) - ig_{aee} a(x) \bar{\psi}(x)\gamma_5\psi(x),
	\label{eq1}
	\end{equation}
where $\hbar$ is the reduced Planck constant, $\gamma^\mu$ is the Dirac matrices vector, $m$ is the mass of the electron and $c$ is the speed of light. The second term of Eq.\,(\ref{eq1}) describes the interaction between $a$ and the spin of the fermion, proportional to the dimensionless coupling constant $g_{aee}$.
In the non-relativistic limit, the interaction term can be expressed as a function of the Bohr magneton $\mu_B$ and of the effective axionic field $\mathbf{B}_a$
	\begin{equation}
	-\frac{g_{aee}\hbar }{2m}\hat{\sigma}\cdot \mathbf{\nabla}a = -2 \frac{e \hbar }{2m}\hat{\sigma}\cdot\left(\frac{g_{aee}}{2e}  \right)\mathbf{\nabla}a \equiv -2\mu_B \hat{\sigma}\cdot \mathbf{B}_a,
	\label{nabla_sigma}
	\end{equation}
where $\hat{\sigma}$ is the Pauli matrices vector and $e$ is the charge of the electron.

Due to the Earth motion through the DM halo of the Galaxy, relic axions can be seen as a wind in an Earth-based laboratory, thus a non zero value of $\nabla a$ is expected. The DM wind average speed is  $v_a\simeq220\,$km/s with a dispersion of about 270\,km/s \cite{PhysRevD.42.1001}.
Axions will interact with an electron spin as an effective magnetic field pointing roughly in the direction of Vega \cite{a50b3ba39b3d4d4b96edd70993223af4,doi:10.1093/mnras/221.4.1023}.
The effective field frequency $f_a$ and amplitude $B_a$ are determined by the mass of the axion $m_a$ and the coupling constant $g_{aee}=3\times10^{-11}(m_a/1\,\mathrm{eV})$
	\begin{align}
	\begin{split}
	\frac{\omega_a}{2\pi}&=f_a=\frac{m_ac^2}{h}\simeq14\,\Big( \frac{m_a}{58.5\,\mu\mathrm{eV}} \Big)\,\mathrm{GHz}, \\
	B_a&=\frac{g_{aee}}{2e}\sqrt{\frac{\hbar n_a}{m_a c}}m_a v_a \\
	&=7\times10^{-23}\,\Big( \frac{\varrho_\mathrm{dm}}{0.45\,\mathrm{GeV}} \Big)^\frac{1}{2}\Big( \frac{m_a}{58.5\,\mu\mathrm{eV}} \Big)\Big( \frac{v_a}{220\,\mathrm{km/s}} \Big)\,\mathrm{T},
	\label{B_values}
	\end{split}
	\end{align}
meaning that $\mathbf{B}_a$ is an extremely weak effective rf magnetic field with a linewidth of $\Delta f_a=7.0\,(m_a/58.5\,\mu\mathrm{eV})\,$kHz, due to the dispersion of $v_a$.
The axion occupation number is $n_a=\varrho_\mathrm{dm}/m_a$, where $\varrho_\mathrm{dm}=0.45$\,GeV/cm$^3$ is the local DM density \cite{pdg}. 
For a reference mass $m_a=58.5\,\mu$eV the mean de Broglie wavelength is $\lambda_{\nabla a}=0.74\lambda_a=0.75h/(m_av_a)=5.1\,\mathrm{m}$, while the coherence time is $\tau_{\nabla a }=0.68\tau_a=58\,\mu$s\footnote{The numerical factors account for the differences between $a$ and $\nabla a$, see \cite{BARBIERI2017135} for further details.}. 

\begin{figure}[h!]
	\centering
	\includegraphics[width=.4\textwidth]{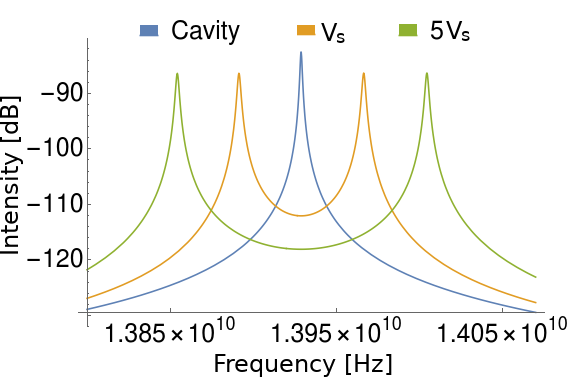}
	\caption{Power spectrum of the cavity (blue line), and hybrid modes calculated for a critically coupled antenna and a sample volume $V_s$ (orange line) and $5V_s$ (green line). The used parameters are close to the experimental values of our apparatus.}
	\label{hyb}
	\end{figure}

Placing a sample in a static magnetic field $\mathbf{B}$, perpendicular to the axion wind, it is possible to tune the Larmor frequency of the electrons to $f_a$, for $\mathbf{B}_0=(0,0,B_0)$, the direction of the electron spin $\hat{\sigma}$ is along the $z$-axis.
The axionic field $\mathbf{B}_a$, acting on the spins of matter, deposits in the material an amount of power $P_\mathrm{in}$
	\begin{equation}
	P_\mathrm{in}=B_a \dv{M}{t} V_s=4\pi \gamma \mu_B f_a B_a^2 \tau_\mathrm{min} n_s V_s,
	\label{pin}
	\end{equation}
where $V_s$ is the volume of the material, $M$ its magnetization, $n_s$ its spin density, $\gamma$ is the gyromagnetic ratio of the electron, and $\tau_\mathrm{min}$ the minimum relaxation time of the system.
The absorbed power is then re-emitted in the form of rf radiation, which can be collected and represents our axion signal.
In free space $\tau_\mathrm{min}$ is mainly determined by radiation damping mechanisms (i.e. magnetic dipole emission of the sample) \cite{AUGUSTINE2002111,doi:10.1063/1.1722859,PhysRev.95.8}, with values much smaller than the material relaxation times.
	\begin{figure}[h!]
	\centering
	\includegraphics[width=.34\textwidth, valign=c]{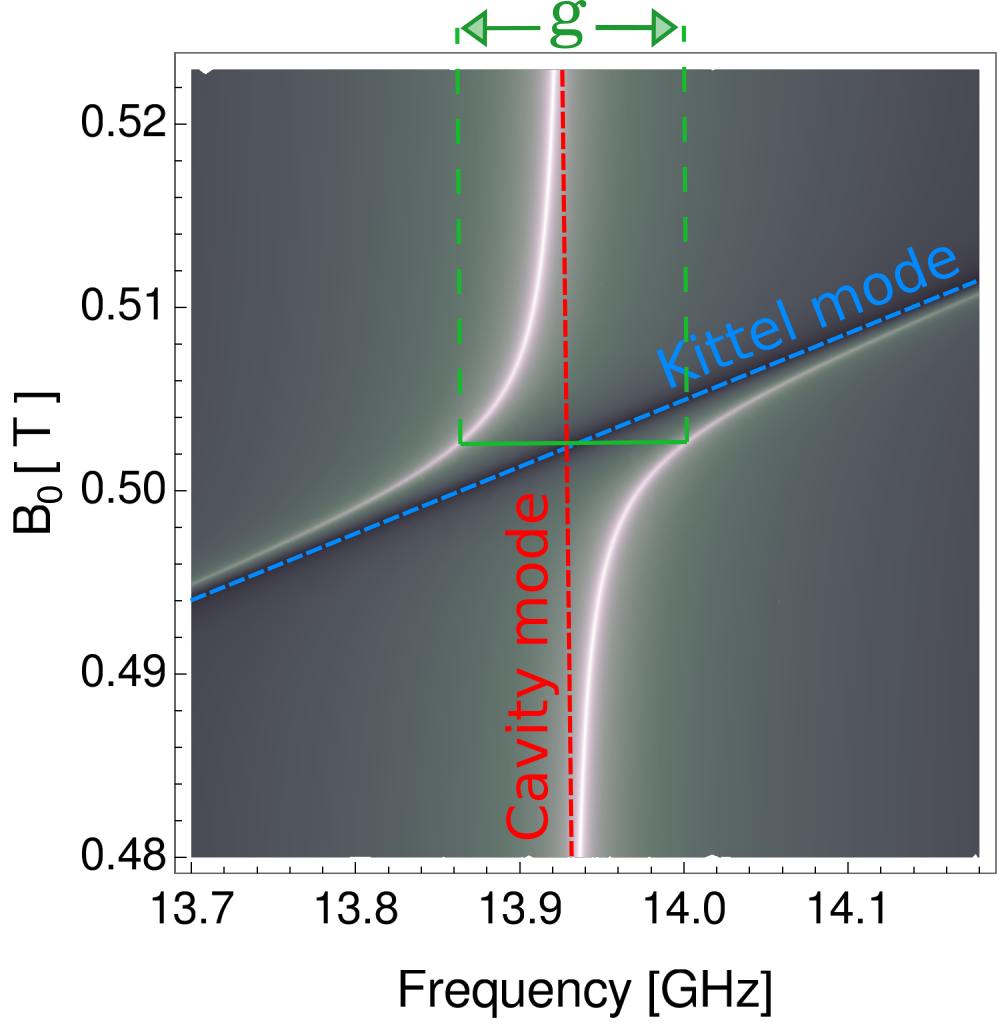}
	\includegraphics[width=0.08\textwidth, valign=c]{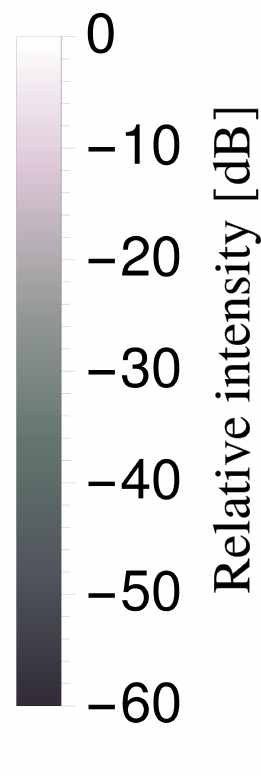}
	\caption{Transmission spectrum of the hybrid system as a function of the external field $B_0$, showing the anticrossing curve of the cavity mode (red dashed line) and Kittel mode (blue dashed line). The coupling $g$ is defined by Eq.\,(\ref{g_coupling}).}
	\label{antic}
	\end{figure}
To avoid the radiation damping issue and thus increase the sensitivity, the magnetic sample is placed inside a resonant cavity. 
A cavity mode with frequency $f_c \simeq f_L$ couples to the Kittel mode (uniform spin precession) of the material, hybridization takes place and the single cavity mode splits into two hybrid modes with frequencies $f_-$ and $f_+$ (strong coupling regime) \cite{kittel,kittel2,Tabuchi405,PhysRevLett.113.083603}. This phenomenon limits the phase space of the dipole emission avoiding radiation damping, and is described by the anti-crossing curve represented in Fig.\,\ref{antic}, which also justifies the strong coupling regime approximation.
The coupling between the cavity mode and the Kittel mode is 
	\begin{equation}
	g=\frac{\gamma}{2\pi}\sqrt{\frac{\mu_0 h f_a}{V_m}n_sV_s}=f_+-f_-,
	\label{g_coupling}
	\end{equation} 
where $\mu_0$ is vacuum magnetic permeability and $V_m=\xi V_c$ is the product of the cavity volume $V_c$ and a mode-dependent form factor $\xi$. The linewidths of the hybrid modes $k_{+,-}$ are an average of the linewidth of the cavity $k_c$ and of the material $k_m$, i.\,e. $k_{+,-}=\frac{1}{2}(k_c+k_m)\equiv k_h$. The calculated power spectral density of an empty cavity and of a cavity with the volume $V_s$ and $5V_s$ of material are shown in Fig.\,\ref{hyb}.
The two hybrid modes are more sensitive to the power deposited by the axion field since they are not affected by radiation damping, the minimum relaxation time is $\tau_\mathrm{min}=\min(\tau_h,\tau_{\nabla a})$, where $\tau_h=1/k_h$.
With an antenna critically coupled to one of the hybrid resonant modes, the extracted power is $P_\mathrm{out}=P_\mathrm{in}/2$.
The scalar product $\hat{\sigma}\cdot\nabla a$ of Eq.\,(\ref{nabla_sigma}) shows that the effect is directional. Due to earth rotation, an earth-based experiment experiences a full daily modulation of the signal, due to the variation of the axion wind direction. 


\section{The QUAX prototype}
\label{quax_prototype}
\begin{figure}[h!]
	\centering
	\includegraphics[width=.45\textwidth]{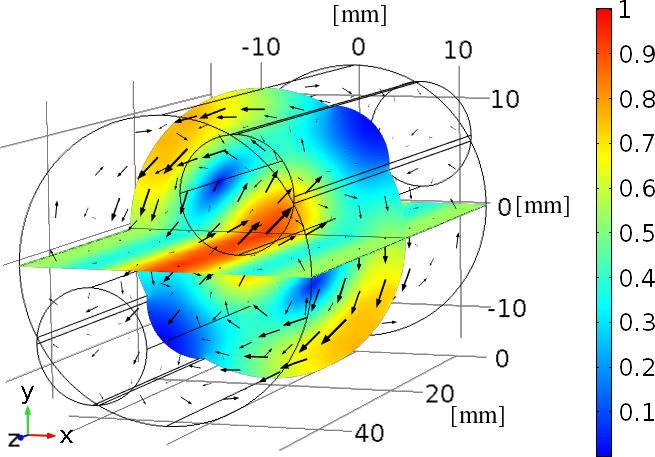}
	\caption{Design of the microwave cavity and magnetic field distribution of the TM110 mode (see text for details). The black arrows represent the direction of the magnetic field, and the color is the normalized field amplitude. The GaYIG spheres are placed on the cavity axis at the maximum of the rf magnetic field.}
	\label{cavity}
	\end{figure}

To implement the scheme presented in Section \ref{esr} we use a cylindrical copper cavity TM110 mode with resonance frequency $f_c\simeq13.98\,$GHz and linewidth $k_c/2\pi\simeq400\,$kHz at liquid helium temperature, measured with a critically coupled antenna. The shape of the cavity is not a regular cylinder, two symmetric sockets are carved into the cylinder to remove the angular degeneration of the normal mode, the maximum and minimum diameters are 26.7\,mm and 26.1\,mm, and the length is 50.0\,mm. The shape of the cavity and of the mode magnetic field are shown in Fig.\,\ref{cavity}. The choice of the TM110 mode has the advantage of having a uniform maximum magnetic rf field along the cavity axis. Its volume can be increased just using a longer cavity without changing the mode resonance frequency. For this mode we calculate a form factor $\xi=0.52$ \cite{doi:10.1063/1.4938164}.
The cavity mode is coupled to a magnetic material, thus we studied the properties of several paramagnetic samples and some ferrites. Highest values of $n_s$ together with long relaxation times have been found for YIG (Yttrium Iron Garnet) and GaYIG (Gallium doped YIG). To avoid inhomogeneous broadening of the linewidth due to geometrical demagnetization, these garnets are shaped as highly polished spheres.
Five GaYIG spheres of 1\,mm diameter have been placed in the maximum magnetic field of the mode, which lies on the axis of the cavity. The spheres are housed inside a PTFE support large enough to let them rotate in all possible directions, in order to automatically align the GaYIG magnetization easy axis with the external magnetic field. 

\begin{figure}[h!]
	\centering
	\includegraphics[width=.21\textwidth]{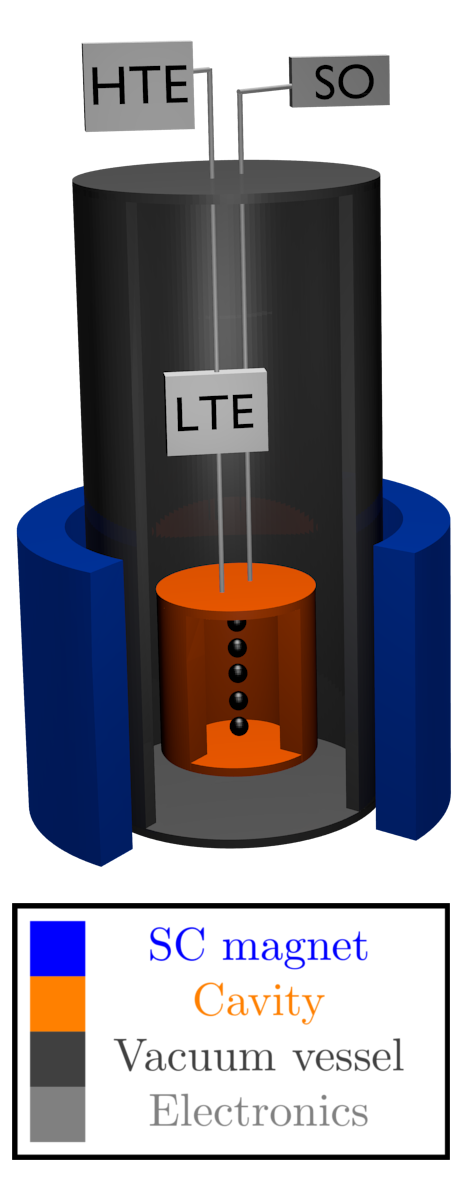}
	\hfill
	\includegraphics[width=.225\textwidth]{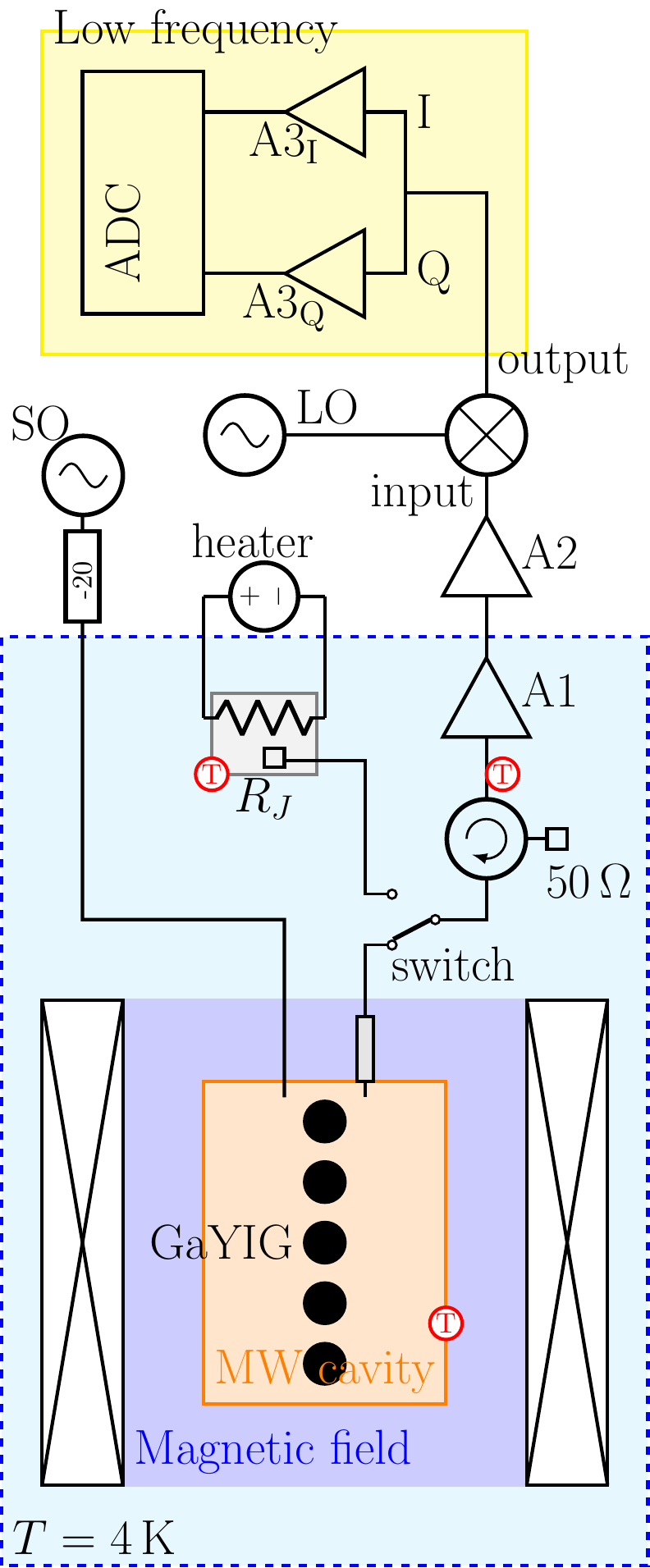}
	\caption{\textit{Left} - Simplified scheme (not to scale) of the experimental apparatus showing the high temperature and low temperature electronics (HTE and LTE) and the source oscillator (SO). \textit{Right} - Electronics layout. From bottom to top, the blue-dashed line encloses the cryogenic part of the apparatus, the crossed rectangles represent the magnet, the orange rectangle is the cavity with black spheres inside standing for the ferrimagnetic material. 
At the top of the cavity are located the sub-critical antenna (left) and the variably-coupled antenna (right). The sub-critical antenna is connected with a room temperature attenuator and then to the source oscillator SO, while the other antenna is connected to one of the switch inputs.
The other input is the 50\,$\Omega$ resistor $R_J$, and the gray rectangle is the plate where $R_J$ is placed and that can be heated with a current generator. The output of the switch is connected to an isolator and then to the A1 and A2 amplifiers. The rf coming from A2 is down-converted by mixing it with a local oscillator LO. The two outputs, phase $I$ and quadrature $Q$, are fed into the low frequency amplifiers A3$_I$ and A3$_Q$, and eventually to the ADC. The red $T$'s are thermometers.}
	\label{app2}
	\end{figure}

The amplitude of an external magnetic field $B_0$ determines the Larmor frequency of the electrons. 
The uniformity of $B_0$ on all the spheres must be enough to avoid inhomogeneous broadening of the ferromagnetic resonance.
To achieve a magnetic field uniformity $\le 1/Q_h$, where $Q_h\sim10^4$ is the quality factor of the hybrid mode, we make use of a superconducting NbTi cylindrical magnet equipped with a concentric cylindrical NbTi correction magnet.
With $B_0=0.5\,$T we have $f_L\simeq f_c$ and thus the hybridization of the cavity and Kittel modes, as discussed in Section \ref{esr}. 
The power supply of the main magnet is a high-precision, high-stability current generator, injecting 15.416\,A into the magnet with a precision better than 1\,mA, while a stable current generator provides 26.0\,A for the correction magnet.
A simplified scheme of the cavity, material and magnet setup is represented in the left part of Fig.\,\ref{app2}.

In the strong coupling regime, the hybrid mode frequencies are $f_+=14.061\,$GHz and $f_-=13.903\,$GHz, yielding a splitting $g=158\,$MHz.
The coupling $g$ scales exactly with $\sqrt{n_sV_s}$, in fact $g=\sqrt{5}\delta$, where $\delta\simeq71\,$MHz is the measured splitting due to a single sphere. This means that all the spins are coherently participating to the material-cavity mode, and ensures that all the spheres magnetization easy axes are aligned along $B_0$.  We use $g$ to calculate the effective number of spins in the sample using the relation described by Eq.\,(\ref{g_coupling}), we obtain $n_s=2.13\times10^{28}\,\mathrm{m}^{-3}$.
The weakly coupled linewidth is $0.7\,$MHz, yielding a critically coupled one of $k_+/2\pi=1.4\,$MHz, corresponding to the hybrid modes relaxation times $\tau_-\simeq\tau_+=0.11\,\mu$s. 

The detection electronics consists in an amplification chain which has two inputs, called Input Channel 1 and 2, (IC-1 and IC-2, respectively). Channel 1 measures the signal power, while Channel 2 has calibration and characterization purposes. A cryogenic switch is used to select the desired channel:
	\begin{description}
	\item[IC-1] - The rf power inside the cavity is collected with a dipole antenna whose coupling to the cavity can be changed using an external micro-manipulator, allowing us to switch continuously from sub-critical to over-critical coupling. For optimal measurement conditions, we tune the antenna to critical coupling by doubling the sub-critical linewidth of the selected mode;
	\item[IC-2] -  A 50\,$\Omega$ termination $R_J$, enclosed in a copper block together with a heater resistance, is used as Johnson noise source. The emitted power can be used to calibrate the noise temperature of the system and the total gain, detailed in Section \ref{calib}.
	\end{description}
The detection electronics, as shown in Fig.\,\ref{app2}, is divided into a liquid helium temperature part (LTE) and a room temperature part(HTE).
The collected power is amplified by a HEMT cryogenic low-noise amplifier (A1) with gain $G_\mathrm{A1}\simeq38\,$dB. To avoid the back-action noise of the amplifier, a cryogenic isolator with 18\,dB of isolation is inserted in the chain. The HTE consists of a room temperature FET amplifier (A2), with $G_\mathrm{A2}\simeq34\,$dB, followed by an IQ mixer used to down-convert the signal with a local oscillator (LO).

The energy distribution of DM axions is highly peaked  ($Q_a \sim 10^6$) around the actual axion mass, so the corresponding frequency distribution can significantly overlap with one (e.\,g. $f_+$) of the two hybrid modes. As the axion mass is unknown, both modes could be monitored independently in order to double  the frequency scan  rate of the detector.  
In our simplified scheme we choose to work only with $f_+$, thus setting the LO frequency to $f_\mathrm{LO}=f_+-0.5\,$MHz and its amplitude to 12\,dBm. The antenna output at the hybrid mode frequency is down-converted in the 0 - 1\,MHz band, allowing us to efficiently digitize the signal. 
The phase and quadrature outputs are fed to two low frequency amplifiers (A3$_{I,Q}$), with a gain of $G_3\simeq50\,$dB each, and are acquired by a 16 bit ADC sampling at 2\,MHz. 

A weakly coupled dipole antenna is used to inject low power signals and make transmission measurements of the system using a source oscillator, SO. All the apparatus devices are referenced to a GPS disciplined, oven controlled, local oscillator.
The cryogenic part of the apparatus is enclosed in a vacuum vessel immersed in liquid helium, as shown schematically in Fig.\,\ref{app2}. Measurements are performed at temperatures $T_c\sim T_a\simeq5.0\,$K and $T_r\simeq5.5\,$K, as read by the cavity, amplifier and $R_J$ thermometers, respectively.

\subsection{Calibration and measurements}
\label{calib}
For the calibration of the system, the load $R_J$ is  heated to a temperature $T_r$, as described in Section \ref{quax_prototype}. Using IC-2 it is possible to measure the Johnson noise of $R_J$ in the temperature range $5 \divisionsymbol 25\,$K without significantly heating other parts of the apparatus.
The rf power from IC-2 is
	\begin{equation}
	P_n=k_B (T_r+T_n) \Delta f,
	\label{pin_cal}
	\end{equation}
where $k_B$ is the Boltzmann constant, $T_n$ is the noise temperature of the system and $\Delta f$ is the bandwidth.
By gradually increasing $T_r$ we linearly change the measured power level of Eq.(\ref{pin_cal}) to determine the noise temperature and gain of the detection electronics, similarly to what is usually obtained with the Y-factor method \cite{BRAGGIO2009451}. 
	\begin{figure}[h!]
	\centering
	\includegraphics[width=.45\textwidth]{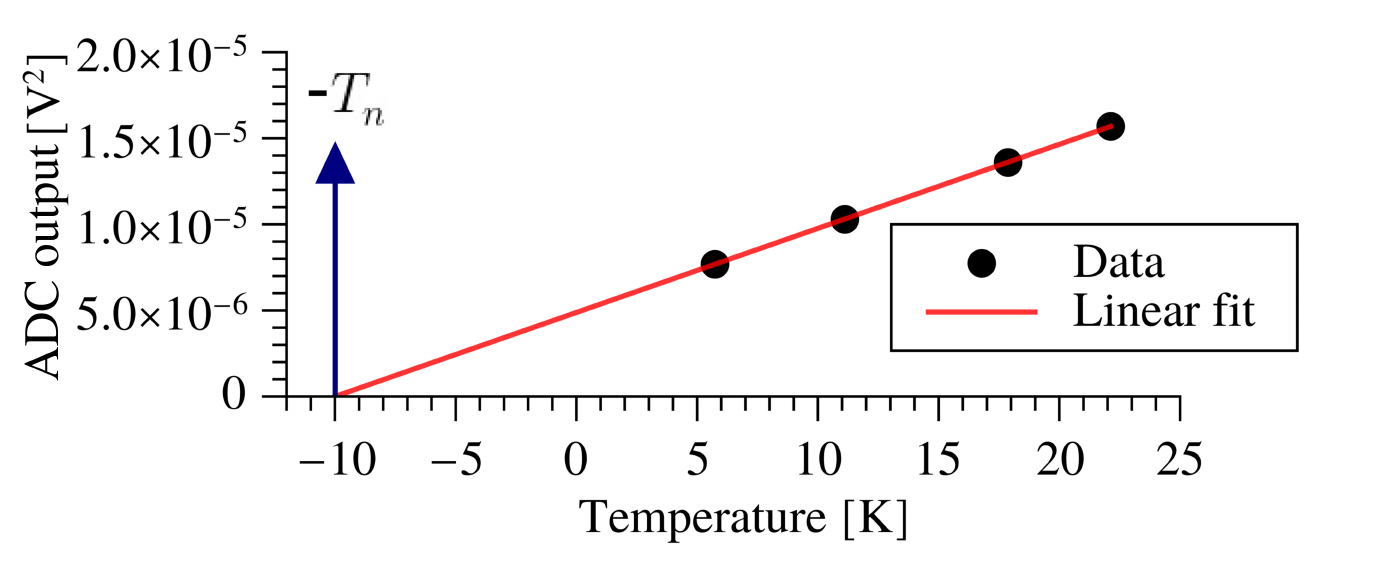}
	\caption{Measurement of noise temperature and gain of the detection electronics.  The statistical error for each point is smaller than the size of the symbol. In the plot the mean square amplitude of the ADC output is plotted vs $T_r$. The noise temperature at the input of A1 is $10^{L/10}\times T_n=8.0\,$K, where $L=-1.0\,$dB are the measured losses between $R_J$ and A1.}
	\label{stab_noiset}
	\end{figure}
In Fig.\,\ref{stab_noiset}, four collected points are fitted with $q(T)=aT+b$ to obtain the noise temperature $T_n=-b/a$ and the total gain $G_\mathrm{tot}=a$.  The error of the estimated parameters is less than 1\%.
Using this method we verified that the noise of the system changes linearly with the temperature, and that the measured cavity output power at the temperature $T_c$ is given by Eq.(\ref{pin_cal}) with $T_r=T_c$, assuming that IC-1 and 2 have the same losses, which is true within 0.2\,dB.
Typical measured values are $T_n=9\divisionsymbol11\,$K and $G_\mathrm{tot}=106\divisionsymbol108\,$dB, at different frequencies around 14\,GHz. This procedure ensures the accuracy of the measurement and then, using IC-1, we perform measurements on the hybrid system with the calibrated electronic chain.

Multiple measurements of the effective axion field have been performed as follows. The vacuum vessel containing the system is cooled down to liquid helium temperature and when a proper thermalization is achieved the detection electronics parameters $G_\mathrm{tot}$ and $T_n$ are measured through IC-2. Then we switch to IC-1, set the magnetic field $B_0$ to $0.5\,$T to hybridize the cavity and Kittel mode at $f_c\simeq f_L\simeq 14\,$GHz, and critically couple the antenna with the $f_+$ hybrid mode using the manipulator. 
A dedicated DAQ software is used to control the oscillators and the ADC, and verifies the correct positioning of the LO with an automated measurement of the hybrid mode transmission spectrum. The ADC digitizes the time-amplitude down-converted signal coming from A3$_I$ and A3$_Q$ and the DAQ software stores collected data binary files of 5\,s each. 
The software also provides a simple online diagnostic, extracting 1\,ms of data every 5\,s, and showing its 512 bin FFT together with the moving average of all FFTs.

As seen in Section \ref{esr}, the axion wind releases a faint power in a band of $\sim7$\,kHz around $f_a$. This signal can be seen only if $f_a$ falls into the detection bandwidth, which corresponds to the linewidth of the hybrid peak.
The expected noise power is given by
\begin{equation}
P_n=1.48\times10^{-18} \Big( \frac{T_c+T_n}{5.2\,\mathrm{K}+10.1\,\mathrm{K}} \Big) \Big( \frac{\Delta f}{7.0\,\mathrm{kHz}} \Big)\,\mathrm{W},
\label{pin_misura}
\end{equation}
calculated from Eq.\,(\ref{pin_cal}) using the data collected from IC-2.
Considering the losses of the system and the gain of the amplifiers,we will show that the mean of the measured power is indeed compatible with the expected noise. 

\subsection{Analysis and results}
\label{sens_res}
The signal is down-converted in its in-phase and quadrature components $\{\phi_n\}$ and $\{q_n\}$, with respect to the local oscillator, that are sampled separately. We applied a complex FFT to $\{s_n\}=\{\phi_n\}+i\{q_n\}$ to get its power spectrum $s^2_\omega$ with positive frequencies for $f>f_\mathrm{LO}$ and negative frequencies for $f<f_\mathrm{LO}$. 
In our experimental settings, the axion signal is mapped almost completely onto the positive frequencies since the hybrid mode linewidth is of order 1\,MHz.

\begin{figure}[h!]
	\centering
	\includegraphics[width=.5\textwidth]{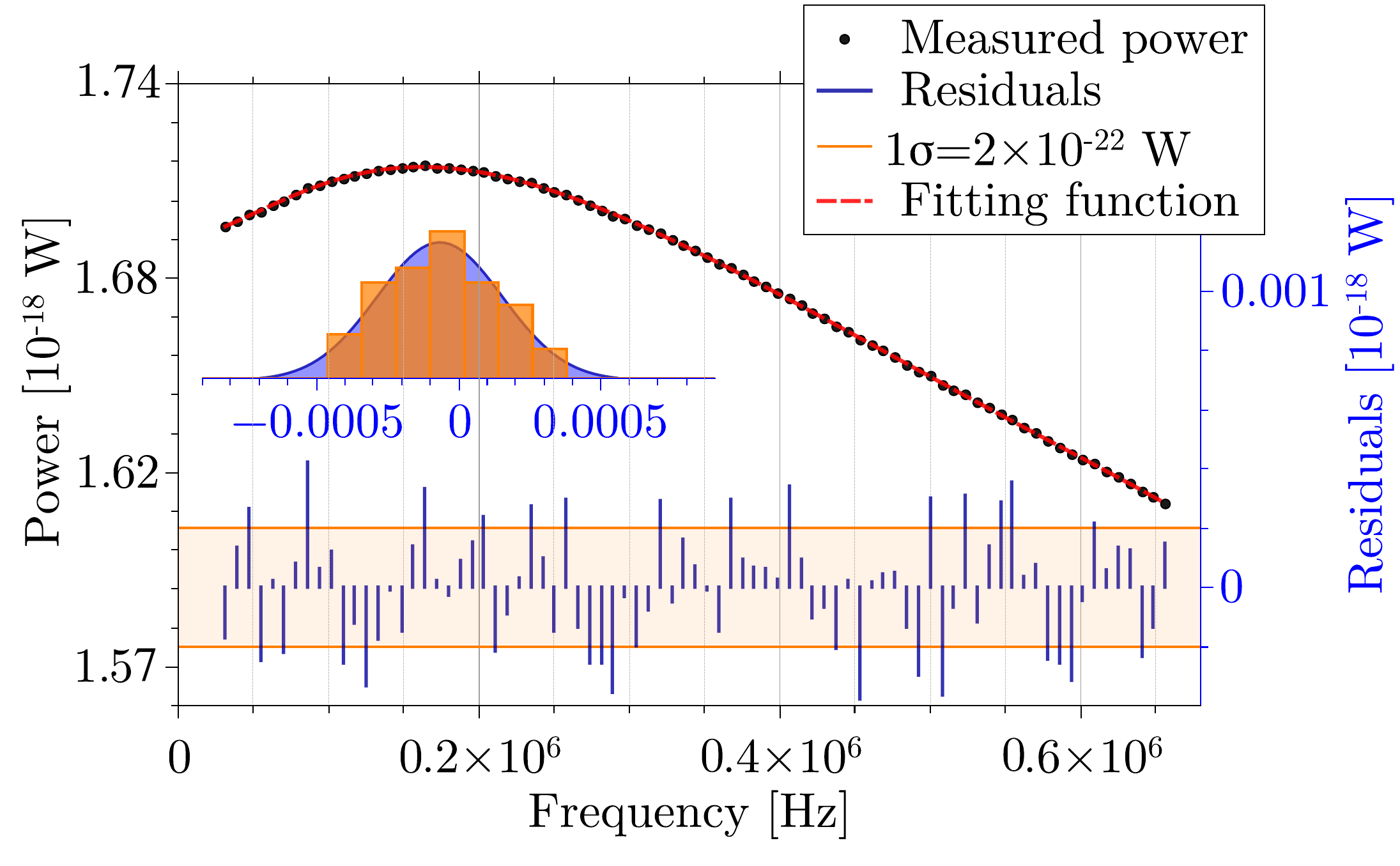}
	\caption{Down-converted power spectrum and residuals of RUN31. The black dots are the measured data points and their error is within the symbol dimensions, the red line is a polynomial fit of such points. The residuals are represented in blue and, as an inset, we show them on an histogram. The corrupted intervals are removed.}
	\label{back}
	\end{figure}

Fig.\,\ref{back} reports the analysis of RUN31, which we describe hereafter in some details.
The $\sim$2.3 hours of the measurement consist in 2048000 FFTs of 8192 bins each (frequency resolution of 244\,Hz), which were square averaged and rebinned to the bandwidth $\Delta f=7.8\,$kHz (256 bins), close to $\Delta f_a$. As explained, we consider only the positive part of the spectrum, consisting of 128 bins, and then calibrate $s^2_\omega$ using Eq.\,(\ref{pin_misura}).
Some frequency intervals of the power spectrum were affected by disturbances at the ADC output, and has been ignored in the analysis procedure. A polynomial of degree 5 is fitted to the averaged spectrum and the residuals estimated. 
The averaged spectrum is reported in Fig.\,\ref{back} together with the fitting function.
In Fig.\,\ref{back} a plot of the residuals and their histogram is also given. The average value of the residuals is $-4.6\times10^{-23}\,$W with standard deviation $\sigma_P=2.2\times10^{-22}\,$W.
The result is compatible with Dicke radiometer equation
	\begin{equation}
	\sigma_\mathrm{D}=k_BT_D\sqrt{\frac{\Delta f}{t}} = 2.1\times10^{-22} \sqrt{\Big(\frac{\Delta f}{7.8\,\mathrm{kHz}} \Big)\Big(\frac{8280\,\mathrm{s}}{t} \Big)}\,\mathrm{W},
	\label{dike}
	\end{equation}
where $t$ is the total integration time and $T_D=T_c+T_n$. This means that the standard deviation of the noise decreases as $1/\sqrt{t}$ trend at least within the RUN31 time span.

\begin{figure}[h!]
	\centering
	\includegraphics[width=.45\textwidth]{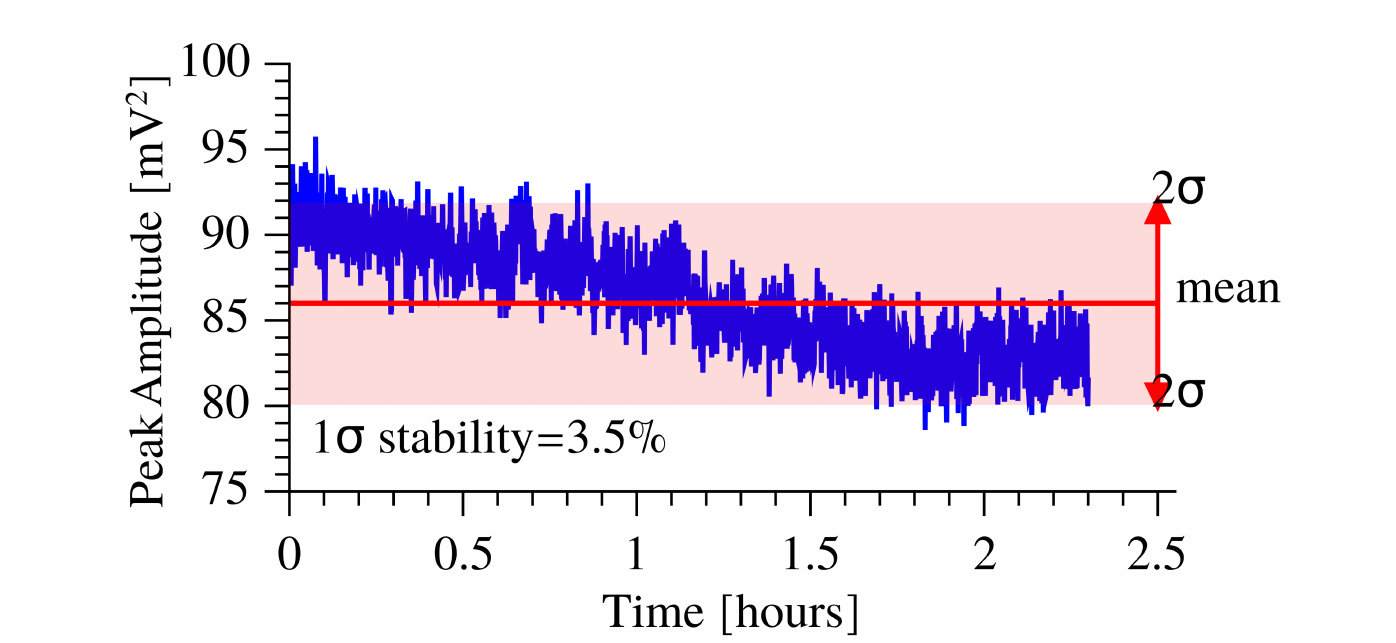}
	\caption{Stability of the hybrid mode, measured through the amplitude of a calibration peak injected with SO (see text for further details).}
	\label{stab_noiset_2}
	\end{figure}

The stability of $f_+$ is monitored by injecting with SO an rf probe signal at $f_+-0.9\,\mathrm{MHz}=f_\mathrm{LO}+0.1\,\mathrm{MHz}$. 
\noindent The transmitted amplitude of the probe peak is a monitor of the hybrid peak frequency since it changes if $f_+$ drifts. Such amplitude is registered during the whole measurement and is plotted in Fig.\,\ref{stab_noiset_2}, the frequency stability of this run was around 3.5\%, which was enough for the purposes of this measurement. We presume that this variation is mostly due to drifts of the external static magnetic field, since with a $B_0=0$ run the corresponding variation was much smaller.
To increase the confidence and the consistency of our estimators, additional offline tests have been performed on the acquired data. Firstly, to search for $P_\mathrm{in}$ when distributed into two adjacent bins, the analysis procedure was repeated using a binning shifted of $\Delta f/2$. This test confirmed the reported result.
Secondly, we calculate the residuals of the averaged spectra for each 5\,s data segment verifying that there are no outliers.  

To infer the axion sensitivity of our measurement, two corrections have to be introduced:
\textit{(i)} a loss of $0.98\,\mathrm{dB}$ (a factor 0.8) at the cavity antenna due to imperfect matching between cavity and axion field \cite{modellorlc};
\textit{(ii)} a factor 1/2 to account for the binning search procedure. 
In fact, the collected power in a single bin results in  $P_\mathrm{in}/2$ because our resolution bandwidth $\Delta f$ overlaps partially  with the axion distribution.
The correct power standard deviation results $\sigma_P^\prime=2\sigma_P/0.8$.

The measured rf power is compatible with the modeled noise for every bin and no statistically significant signal consistent with axions was found. The upper limit at the 95\% C.L. is $2\sigma^\prime_P=1.1\times10^{-21}\,$W. 
This value can be converted to equivalent axion field with the help of Eq.\,(\ref{pin}), obtaining
	\begin{align}
	\begin{split}
	B_m<& \Big( \frac{P_\mathrm{out}=2\sigma_P^\prime}{4\pi \gamma \mu_B n_S f_+ \tau_+ V_s}\Big)^{1/2}=2.6\times10^{-17} \Big[\Big( \frac{14\,\mathrm{GHz}}{f_+} \Big) \times\\
	& \times\Big( \frac{2.13\cdot10^{28}/\mathrm{m}^3}{n_S} \Big) \Big( \frac{0.11\,\mu\mathrm{s}}{\tau_+} \Big) \Big( \frac{2.6\,\mathrm{mm}^3}{V_s} \Big)\Big]^{1/2}\,\mathrm{T},
	\label{field_limit}
	\end{split}
	\end{align}
where all the reported parameters have been explicitly measured.
The limit holds for the central frequency of the hybrid mode, while for other frequencies the sensitivity have to be normalized:
the correct sensitivity is obtained dividing $B_m$ by the normalized amplitude of the hybrid mode Lorentzian.
Several measurements have been performed for different cool downs of the setup. Probably due to mechanical instabilities and to the low resolution of the correction magnet power supply, the resulting working frequency $f_+$ slightly changed between the runs, allowing us to perform also a limited frequency scan over a $\sim$3\,MHz range.  The maximum integration time for a 1\,MHz band was 6 hours, and no deviations from the $1/\sqrt{t}$ scaling of $\sigma_P$ were found. 

\section{Discussions}
\label{discussioni}
Our results represent also a limit on the axion-electron coupling constant. Since $B_m$ depends on $g_{aee}$, the explicit form of the effective magnetic field given in Eq.\,(\ref{B_values}) can be recast with the help of Eq.\,(\ref{pin}), to
	\begin{equation}
	g_{aee}>\frac{e}{\pi m_av_a}\sqrt{\frac{2\sigma^\prime_P}{2\mu_B \gamma\, n_a n_s V_s \tau_+}},
	\label{gaee2}
	\end{equation}
at 95\% confidence level.
The results of this preliminary measurements are far from the sensitivity requirements for a cosmological axion search [see Eq.\,(\ref{B_values})], however they can be used to detect DM Axion-like particles (ALPs), which can account for the whole dark matter density \cite{1475-7516-2012-06-013}.
During the measurement time the DM-wind amplitude was on the maximum of the daily modulation, allowing us to use the collected data to obtain an upper limit on the ALP-electron coupling at the maximum sensitivity.
Through Eq.\,(\ref{gaee2}) we are able to exclude values of the ALP-electron coupling constant for ALP masses given by $f_+$ through Eq.\,(\ref{B_values}).

\begin{figure}
	\includegraphics[width=.5\textwidth]{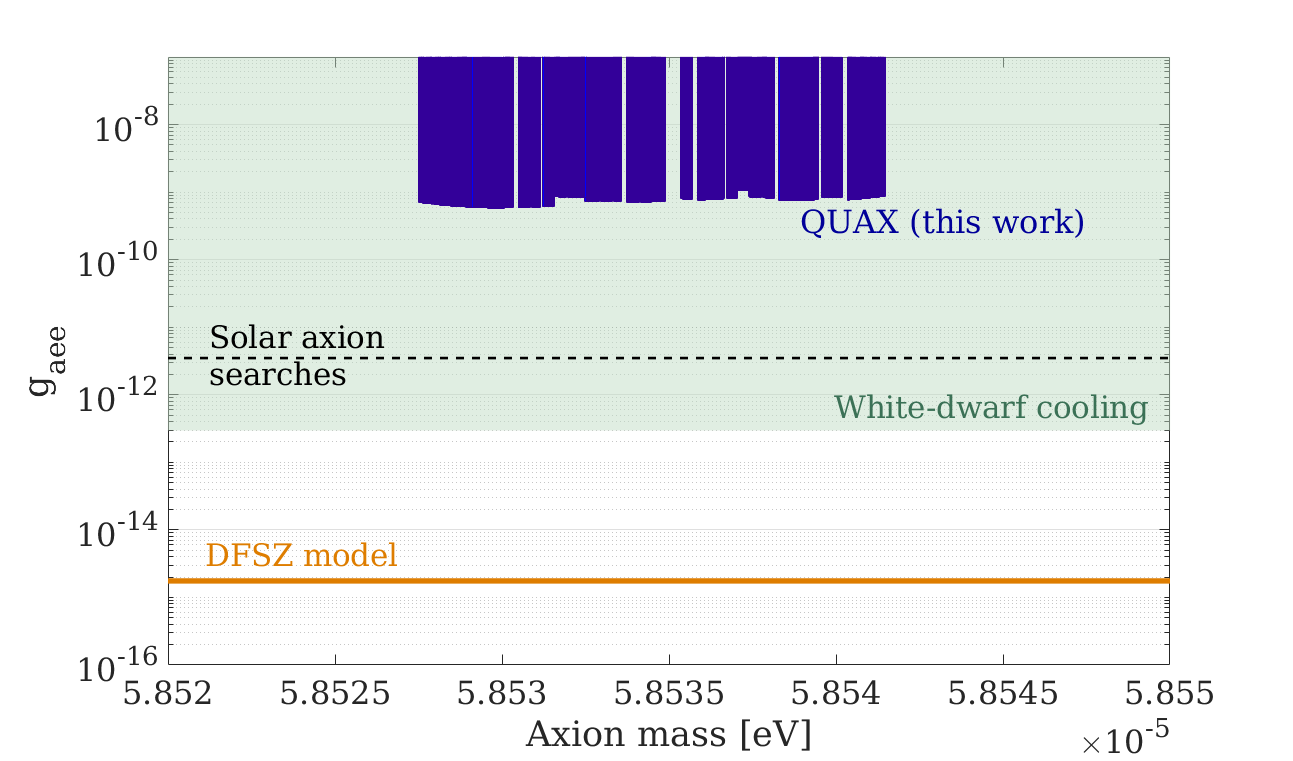}
	\caption{Excluded values of the $g_{aee}$ coupling (blue area) compared to its theoretical prediction for the DFSZ axion model with $\beta=1$ (orange line) and a DM density of 0.45\,GeV/cm$^3$. The green shaded area is excluded by white dwarf cooling \cite{raffeltaxions,battich,Corsico:2016okh}, while the black dashed line is the best upper limit obtained with solar axion searches relying on the axio-electric effect \cite{PhysRevLett.118.261301,201346,PhysRevD.90.062009,PhysRevD.95.029904,1475-7516-2013-11-067,Yoon2016,Derbin2012}. Other statistically significant limits can be found in \cite{1475-7516-2017-10-010,cicoli}.}
	\label{limit}
	\end{figure}

By repeating the analysis procedure described in Section \ref{sens_res} for seven measurement runs and averaging together overlapping bandwidths, we produce the plot in Fig.\,\ref{limit}. 
The minimum measured value of $g_{aee}$ is $4.9\times10^{-10}$, corresponding to an equivalent axion field limit of $1.6\times 10^{-17}\,$T.

\subsection{Improvements and discovery potential}
\label{improvements}
To push the present sensitivity towards smaller values of the coupling constant $g_{aee}$, several improvements should be implemented.
In fact, using Eq.\,(\ref{pin}), the power released by a DFSZ-axion wind in the five GaYIG spheres of our prototype is
	\begin{align}
	\begin{split}
	P_{\rm out}=&\frac{P_{\rm in}}{2}=1.4 \times 10^{-33} \left(\frac{m_a}{58.5\, \mu{\rm eV}}\right)^3 \times\\
	&\times \left( \frac{n_s}{2 \cdot10^{28} /{\rm m}^3}\right)\left( \frac{V_s}{2.6\,\, {\rm mm}^3}\right)
	 \left( \frac{\tau_{\rm min}}{0.11 \,\mu{\rm s}}\right)\,{\rm W},
	\label{power}
	\end{split}
	\end{align}
corresponding to a rate $r_a\sim10^{-10}\,\mathrm{Hz}$ of 14\,GHz photons, which is clearly not detectable.
To have a statistically significant signal within a reasonable integration time it is mandatory to increase the signal rate, for example in the mHz range, that will give tens counts per day.
The present sensitivity to the power deposited in the system by the axion wind maintains an excess photon rate of order $100\,$photons/s. 

Short term improvements that will be installed in the prototype include a larger volume of narrow-linewidth magnetic material, namely 10 YIG spheres of 2\,mm diameter, a lower cavity temperature with dilution refrigeration and the use of a Josephson Parametric Amplifier (JPA).
The increased volume will enhance the axionic signal of a factor 16. As for the noise reduction, a working temperature of 100\,mK will reduce the thermal fluctuations and there are hints suggesting that it can also reduce the YIG linewidth. Ultra cryogenic temperatures allow us to use JPAs as first-stage amplifier to drastically increase the sensitivity, since its noise temperature can be of the order of 100\,mK. The upgraded prototype should be capable of setting a limit on the effective magnetic field $B_m$ two orders of magnitude better than the present one.

To achieve the QUAX goal \cite{BARBIERI2017135}, the detector requires an improvement of three to four more orders of magnitude in sensitivity, which can be obtained increasing the signal and reducing the noise. Using a $V_s\simeq0.1\,$liters and $\tau_\mathrm{min}\simeq1\,\mu$s, the axionic power deposited in the system is $\sim 10^{-27}\,$W. This power is smaller than the quantum noise, meaning that a quantum counter, immune to such noise, must be exploited to push the sensitivity to the axion level \cite{PhysRevD.88.035020,kuzmin}.

To scan different axion masses we must vary the working frequency of the haloscope. A large tuning can be achieved by changing both the cavity mode resonance frequency and the Larmor frequency (i.\,e. the static magnetic field $B_0$). A small frequency tuning is possible by varying only $B_0$: in this case, a scanning of several MHz is possible without a significant reduction of the sensitivity.

In the favored case of a signal detection, its nature can be systematically studied by QUAX. Since the axion signal is persistent, it will be possible to infer DM properties by using the directionality of the apparatus. Moreover, this setup is able to test different axion models, measuring separately the axion-to-photon and axion-to-electron couplings. In fact the apparatus has also the features of a Sikivie haloscope \cite{PhysRevLett.51.1415}, and can be sensitive to the axion-photon coupling by using a suitable cavity mode.

\section{Conclusions}
\label{conclusion}
We described the operation of a prototype of the QUAX experiment, a ferromagnetic haloscope sensitive to DM axion through their interaction with electron spin. Our findings indicate the possibility of performing electron spin resonance measurements of a sizable quantity of material inside a cavity cooled down to cryogenic temperatures. By using low noise electronics we search for extra power injected in the system that could be due to DM axions. We reach a power sensitivity of $10^{-22}\,$W that can be translated to an upper limit on the the coupling constant $g_{aee}<4.9\times10^{-10}$ for an axion mass of $58\,\mu$eV, which, to our knowledge, is the first measurement of the coupling between cosmological axions and electrons. The sensitivity of our apparatus is presently limited only by the noise temperature of the system and thermodynamic fluctuations, as it reaches the limit of Dicke radiometer equation. The overall behavior of the apparatus is as expected, and thus we are confident that the planned upgrades will be effective.

\section*{Acknowledgments}
\label{ack}
The authors wish to thank Fulvio Calaon, Mario Tessaro, Mario Zago, Massimo Rebeschini, Andrea Benato and Enrico Berto for the help with cryogenics and for the mechanical and electronic work on the experimental setup. We acknowledge the support of Giampaolo Galet and Lorenzo Castellani for the building of the high-precision power supply and Nicola Toniolo, Michele Gulmini and Stefano Marchini for the work on the DAQ system. We also acknowledge Paolo Falferi and Renato Mezzena for the useful experimental suggestions, and finally we thank Riccardo Barbieri and Andreas Ringwald for the stimulating theoretical discussion and advice.

\bibliographystyle{spphys}
\bibliography{quax_gaee}

\begin{thebibliography}{10}
\providecommand{\url}[1]{{#1}}
\providecommand{\urlprefix}{URL }
\expandafter\ifx\csname urlstyle\endcsname\relax
  \providecommand{\doi}[1]{DOI \discretionary{}{}{}#1}\else
  \providecommand{\doi}{DOI \discretionary{}{}{}\begingroup
  \urlstyle{rm}\Url}\fi

\bibitem{zwicky}
F.~{Zwicky}, Helvetica Physica Acta \textbf{6}, 110 (1933)

\bibitem{rubin78}
V.C. {Rubin}, N.~{Thonnard}, W.K. {Ford}, Jr., Astrophysical Journal
  \textbf{225}, L107 (1978).
\newblock \doi{10.1086/182804}

\bibitem{rubin80}
V.C. Rubin, W.K. Ford, Jr., N.~Thonnard, Astrophysical Journal \textbf{238},
  471 (1980).
\newblock \doi{10.1086/158003}

\bibitem{pdg}
C.~Patrignani, et~al., Chin. Phys. \textbf{C40}(10), 100001 (2016).
\newblock \doi{10.1088/1674-1137/40/10/100001}

\bibitem{pq}
R.D. Peccei, H.R. Quinn, Phys. Rev. Lett. \textbf{38}, 1440 (1977).
\newblock \doi{10.1103/PhysRevLett.38.1440}.
\newblock \urlprefix\url{http://link.aps.org/doi/10.1103/PhysRevLett.38.1440}

\bibitem{weinberg1978new}
S.~Weinberg, Physical Review Letters \textbf{40}(4), 223 (1978)

\bibitem{DINE1981199}
M.~Dine, W.~Fischler, M.~Srednicki, Physics Letters B \textbf{104}(3), 199
  (1981).
\newblock \doi{https://doi.org/10.1016/0370-2693(81)90590-6}.
\newblock
  \urlprefix\url{http://www.sciencedirect.com/science/article/pii/0370269381905906}

\bibitem{SHIFMAN1980493}
M.~Shifman, A.~Vainshtein, V.~Zakharov, Nuclear Physics B \textbf{166}(3), 493
  (1980).
\newblock \doi{https://doi.org/10.1016/0550-3213(80)90209-6}.
\newblock
  \urlprefix\url{http://www.sciencedirect.com/science/article/pii/0550321380902096}

\bibitem{PhysRevLett.43.103}
J.E. Kim, Phys. Rev. Lett. \textbf{43}, 103 (1979).
\newblock \doi{10.1103/PhysRevLett.43.103}.
\newblock \urlprefix\url{https://link.aps.org/doi/10.1103/PhysRevLett.43.103}

\bibitem{DINE1983137}
M.~Dine, W.~Fischler, Physics Letters B \textbf{120}(1), 137  (1983).
\newblock \doi{https://doi.org/10.1016/0370-2693(83)90639-1}.
\newblock
  \urlprefix\url{http://www.sciencedirect.com/science/article/pii/0370269383906391}

\bibitem{wilczek1978problem}
F.~Wilczek, Physical Review Letters \textbf{40}(5), 279 (1978)

\bibitem{PRESKILL1983127}
J.~Preskill, M.B. Wise, F.~Wilczek, Physics Letters B \textbf{120}(1), 127
  (1983).
\newblock \doi{https://doi.org/10.1016/0370-2693(83)90637-8}.
\newblock
  \urlprefix\url{http://www.sciencedirect.com/science/article/pii/0370269383906378}

\bibitem{raffelt1996stars}
G.G. Raffelt, \emph{Stars as laboratories for fundamental physics: The
  astrophysics of neutrinos, axions, and other weakly interacting particles}
  (University of Chicago press, 1996)

\bibitem{turner1990windows}
M.S. Turner, Physics Reports \textbf{197}(2), 67 (1990)

\bibitem{abbott1983cosmological}
L.F. Abbott, P.~Sikivie, Physics Letters B \textbf{120}(1-3), 133 (1983)

\bibitem{fox2004probing}
P.~Fox, A.~Pierce, S.~Thomas, arXiv preprint hep-th/0409059  (2004)

\bibitem{spergel170wilkinson}
D.~Spergel, et~al., Astrophys. J. Suppl \textbf{170}, 377

\bibitem{BURGER2017880}
F.~Burger, E.M. Ilgenfritz, M.P. Lombardo, M.~Müller-Preussker, A.~Trunin,
  Nuclear Physics A \textbf{967}, 880  (2017).
\newblock \doi{https://doi.org/10.1016/j.nuclphysa.2017.07.006}.
\newblock
  \urlprefix\url{http://www.sciencedirect.com/science/article/pii/S037594741730324X}.
\newblock The 26th International Conference on Ultra-relativistic
  Nucleus-Nucleus Collisions: Quark Matter 2017

\bibitem{berkowitz2015lattice}
E.~Berkowitz, M.I. Buchoff, E.~Rinaldi, Physical Review D \textbf{92}(3),
  034507 (2015)

\bibitem{borsanyi2016calculation}
S.~Bors{\'a}nyi, Z.~Fodor, J.~Guenther, et~al., Nature \textbf{539}(7627), 69
  (2016)

\bibitem{diCortona2016}
G.G. di~Cortona, E.~Hardy, J.P. Vega, G.~Villadoro, Journal of High Energy
  Physics \textbf{2016}(1), 34 (2016).
\newblock \urlprefix\url{https://doi.org/10.1007/JHEP01(2016)034}

\bibitem{petreczky2016topological}
P.~Petreczky, H.P. Schadler, S.~Sharma, Physics Letters B \textbf{762}, 498
  (2016)

\bibitem{bonati}
{Bonati, Claudio}, {D’Elia, Massimo}, {Mariti, Marco}, et~al., EPJ Web Conf.
  \textbf{137}, 08004 (2017).
\newblock \doi{10.1051/epjconf/201713708004}.
\newblock \urlprefix\url{https://doi.org/10.1051/epjconf/201713708004}

\bibitem{PhysRevLett.51.1415}
P.~Sikivie, Phys. Rev. Lett. \textbf{51}, 1415 (1983).
\newblock \doi{10.1103/PhysRevLett.51.1415}.
\newblock \urlprefix\url{https://link.aps.org/doi/10.1103/PhysRevLett.51.1415}

\bibitem{ringwald}
J.~Jaeckel, A.~Ringwald, Annual Review of Nuclear and Particle Science
  \textbf{60}(1), 405 (2010).
\newblock \doi{10.1146/annurev.nucl.012809.104433}.
\newblock \urlprefix\url{https://doi.org/10.1146/annurev.nucl.012809.104433}

\bibitem{axion_searches}
P.W. Graham, I.G. Irastorza, S.K. Lamoreaux, A.~Lindner, K.A. van Bibber,
  Annual Review of Nuclear and Particle Science \textbf{65}(1), 485 (2015).
\newblock \doi{10.1146/annurev-nucl-102014-022120}.
\newblock \urlprefix\url{https://doi.org/10.1146/annurev-nucl-102014-022120}

\bibitem{redondo}
I.G. Irastorza, J.~Redondo,   (2018).
\newblock \urlprefix\url{https://arxiv.org/abs/1801.08127}

\bibitem{kim}
J.E. Kim, AIP Conference Proceedings \textbf{1200}(1), 83 (2010).
\newblock \doi{10.1063/1.3327743}.
\newblock \urlprefix\url{https://aip.scitation.org/doi/abs/10.1063/1.3327743}

\bibitem{PhysRevLett.112.091302}
M.~Arik, S.~Aune, K.~Barth, et~al., Phys. Rev. Lett. \textbf{112}, 091302
  (2014).
\newblock \doi{10.1103/PhysRevLett.112.091302}.
\newblock
  \urlprefix\url{https://link.aps.org/doi/10.1103/PhysRevLett.112.091302}

\bibitem{1475-7516-2007-04-010}
S.~Andriamonje, S.~Aune, D.~Autiero, et~al., Journal of Cosmology and
  Astroparticle Physics \textbf{2007}(04), 010 (2007).
\newblock \urlprefix\url{http://stacks.iop.org/1475-7516/2007/i=04/a=010}

\bibitem{PhysRevLett.118.061302}
B.M. Brubaker, L.~Zhong, Y.V. Gurevich, et~al., Phys. Rev. Lett. \textbf{118},
  061302 (2017).
\newblock \doi{10.1103/PhysRevLett.118.061302}.
\newblock
  \urlprefix\url{https://link.aps.org/doi/10.1103/PhysRevLett.118.061302}

\bibitem{PhysRevLett.104.041301}
S.J. Asztalos, G.~Carosi, C.~Hagmann, et~al., Phys. Rev. Lett. \textbf{104},
  041301 (2010).
\newblock \doi{10.1103/PhysRevLett.104.041301}.
\newblock
  \urlprefix\url{https://link.aps.org/doi/10.1103/PhysRevLett.104.041301}

\bibitem{PhysRevD.74.012006}
L.D. Duffy, P.~Sikivie, D.B. Tanner, et~al., Phys. Rev. D \textbf{74}, 012006
  (2006).
\newblock \doi{10.1103/PhysRevD.74.012006}.
\newblock \urlprefix\url{https://link.aps.org/doi/10.1103/PhysRevD.74.012006}

\bibitem{MCALLISTER201767}
B.T. McAllister, G.~Flower, E.N. Ivanov, et~al., Physics of the Dark Universe
  \textbf{18}, 67  (2017).
\newblock \doi{https://doi.org/10.1016/j.dark.2017.09.010}.
\newblock
  \urlprefix\url{http://www.sciencedirect.com/science/article/pii/S2212686417300602}

\bibitem{refId0}
{Petrakou, Eleni}, {for CAPP/IBS}, EPJ Web Conf. \textbf{164}, 01012 (2017).
\newblock \doi{10.1051/epjconf/201716401012}.
\newblock \urlprefix\url{https://doi.org/10.1051/epjconf/201716401012}

\bibitem{caldwell2017dielectric}
A.~Caldwell, G.~Dvali, B.~Majorovits, et~al., Physical review letters
  \textbf{118}(9), 091801 (2017)

\bibitem{PhysRevLett.120.151301}
N.~Du, N.~Force, R.~Khatiwada, et~al., Phys. Rev. Lett. \textbf{120}, 151301
  (2018).
\newblock \doi{10.1103/PhysRevLett.120.151301}.
\newblock
  \urlprefix\url{https://link.aps.org/doi/10.1103/PhysRevLett.120.151301}

\bibitem{PhysRevLett.118.071802}
G.~Ballesteros, J.~Redondo, A.~Ringwald, C.~Tamarit, Phys. Rev. Lett.
  \textbf{118}, 071802 (2017).
\newblock \doi{10.1103/PhysRevLett.118.071802}.
\newblock
  \urlprefix\url{https://link.aps.org/doi/10.1103/PhysRevLett.118.071802}

\bibitem{1475-7516-2017-08-001}
G.~Ballesteros, J.~Redondo, A.~Ringwald, C.~Tamarit, Journal of Cosmology and
  Astroparticle Physics \textbf{2017}(08), 001 (2017).
\newblock \urlprefix\url{http://stacks.iop.org/1475-7516/2017/i=08/a=001}

\bibitem{Ernst2018}
A.~Ernst, A.~Ringwald, C.~Tamarit, Journal of High Energy Physics
  \textbf{2018}(2), 103 (2018).
\newblock \doi{10.1007/JHEP02(2018)103}.
\newblock \urlprefix\url{https://doi.org/10.1007/JHEP02(2018)103}

\bibitem{Ema2017}
Y.~Ema, K.~Hamaguchi, T.~Moroi, K.~Nakayama, Journal of High Energy Physics
  \textbf{2017}(1), 96 (2017).
\newblock \doi{10.1007/JHEP01(2017)096}.
\newblock \urlprefix\url{https://doi.org/10.1007/JHEP01(2017)096}

\bibitem{PhysRevD.95.095009}
L.~Calibbi, F.~Goertz, D.~Redigolo, R.~Ziegler, J.~Zupan, Phys. Rev. D
  \textbf{95}, 095009 (2017).
\newblock \doi{10.1103/PhysRevD.95.095009}.
\newblock \urlprefix\url{https://link.aps.org/doi/10.1103/PhysRevD.95.095009}

\bibitem{Garcon:2017ixh}
A.~Garcon, D.~Aybas, J.W. Blanchard, et~al., Quantum Science and Technology
  \textbf{3}(1), 014008 (2018).
\newblock \urlprefix\url{http://stacks.iop.org/2058-9565/3/i=1/a=014008}

\bibitem{PhysRevX.4.021030}
D.~Budker, P.W. Graham, M.~Ledbetter, S.~Rajendran, A.O. Sushkov, Phys. Rev. X
  \textbf{4}, 021030 (2014).
\newblock \doi{10.1103/PhysRevX.4.021030}.
\newblock \urlprefix\url{https://link.aps.org/doi/10.1103/PhysRevX.4.021030}

\bibitem{1742-6596-718-4-042051}
G.~Ruoso, A.~Lombardi, A.~Ortolan, et~al., Journal of Physics: Conference
  Series \textbf{718}(4), 042051 (2016).
\newblock \urlprefix\url{http://stacks.iop.org/1742-6596/718/i=4/a=042051}

\bibitem{BARBIERI2017135}
R.~Barbieri, C.~Braggio, G.~Carugno, et~al., Physics of the Dark Universe
  \textbf{15}, 135  (2017).
\newblock \doi{https://doi.org/10.1016/j.dark.2017.01.003}.
\newblock
  \urlprefix\url{http://www.sciencedirect.com/science/article/pii/S2212686417300031}

\bibitem{vorobyov1995ferromagnetic}
P.~Vorobyov, A.~Kirpotin, M.~Rovkin, A.~Boldyrev, arXiv preprint hep-ph/9506371
   (1995)

\bibitem{BARBIERI1989357}
R.~Barbieri, M.~Cerdonio, G.~Fiorentini, S.~Vitale, Physics Letters B
  \textbf{226}(3), 357  (1989).
\newblock \doi{https://doi.org/10.1016/0370-2693(89)91209-4}.
\newblock
  \urlprefix\url{http://www.sciencedirect.com/science/article/pii/0370269389912094}

\bibitem{kakhidze1991antiferromagnetic}
A.~Kakhidze, I.~Kolokolov, Zh. Eksp. Teor. Fiz \textbf{99}, 1077 (1991)

\bibitem{Caspers:1989ix}
F.~Caspers, Y.~Semertzidis, in \emph{{Cosmic axions. Proceedings, Workshop,
  Upton, USA, April 13-14, 1989}} (1989), pp. 0173--183

\bibitem{PhysRevD.42.1001}
M.S. Turner, F.~Wilczek, Phys. Rev. D \textbf{42}, 1001 (1990).
\newblock \doi{10.1103/PhysRevD.42.1001}.
\newblock \urlprefix\url{https://link.aps.org/doi/10.1103/PhysRevD.42.1001}

\bibitem{a50b3ba39b3d4d4b96edd70993223af4}
P.J. McMillan, J.J. Binney, Monthly Notices of the Royal Astronomical Society
  \textbf{402}, 934 (2010).
\newblock \doi{10.1111/j.1365-2966.2009.15932.x}

\bibitem{doi:10.1093/mnras/221.4.1023}
F.J. Kerr, D.~Lynden-Bell, Monthly Notices of the Royal Astronomical Society
  \textbf{221}(4), 1023 (1986).
\newblock \doi{10.1093/mnras/221.4.1023}.
\newblock \urlprefix\url{http://dx.doi.org/10.1093/mnras/221.4.1023}

\bibitem{AUGUSTINE2002111}
M.~Augustine, Progress in Nuclear Magnetic Resonance Spectroscopy
  \textbf{40}(2), 111  (2002).
\newblock \doi{https://doi.org/10.1016/S0079-6565(01)00037-1}.
\newblock
  \urlprefix\url{http://www.sciencedirect.com/science/article/pii/S0079656501000371}

\bibitem{doi:10.1063/1.1722859}
S.~Bloom, Journal of Applied Physics \textbf{28}(7), 800 (1957).
\newblock \doi{10.1063/1.1722859}

\bibitem{PhysRev.95.8}
N.~Bloembergen, R.V. Pound, Phys. Rev. \textbf{95}, 8 (1954).
\newblock \doi{10.1103/PhysRev.95.8}.
\newblock \urlprefix\url{https://link.aps.org/doi/10.1103/PhysRev.95.8}

\bibitem{kittel}
C.~Kittel, Phys. Rev. \textbf{71}, 270 (1947).
\newblock \doi{10.1103/PhysRev.71.270.2}.
\newblock \urlprefix\url{https://link.aps.org/doi/10.1103/PhysRev.71.270.2}

\bibitem{kittel2}
M.~Sparks, C.~Kittel, Phys. Rev. Lett. \textbf{4}, 232 (1960).
\newblock \doi{10.1103/PhysRevLett.4.232}.
\newblock \urlprefix\url{https://link.aps.org/doi/10.1103/PhysRevLett.4.232}

\bibitem{Tabuchi405}
Y.~Tabuchi, S.~Ishino, A.~Noguchi, et~al., Science \textbf{349}(6246), 405
  (2015).
\newblock \doi{10.1126/science.aaa3693}.
\newblock \urlprefix\url{http://science.sciencemag.org/content/349/6246/405}

\bibitem{PhysRevLett.113.083603}
Y.~Tabuchi, S.~Ishino, T.~Ishikawa, et~al., Phys. Rev. Lett. \textbf{113},
  083603 (2014).
\newblock \doi{10.1103/PhysRevLett.113.083603}.
\newblock
  \urlprefix\url{https://link.aps.org/doi/10.1103/PhysRevLett.113.083603}

\bibitem{doi:10.1063/1.4938164}
I.~Stern, A.A. Chisholm, J.~Hoskins, et~al., Review of Scientific Instruments
  \textbf{86}(12), 123305 (2015).
\newblock \doi{10.1063/1.4938164}.
\newblock \urlprefix\url{https://doi.org/10.1063/1.4938164}

\bibitem{BRAGGIO2009451}
C.~Braggio, G.~Bressi, G.~Carugno, et~al., Nuclear Instruments and Methods in
  Physics Research Section A: Accelerators, Spectrometers, Detectors and
  Associated Equipment \textbf{603}(3), 451  (2009).
\newblock \doi{https://doi.org/10.1016/j.nima.2009.02.021}.
\newblock
  \urlprefix\url{http://www.sciencedirect.com/science/article/pii/S0168900209003921}

\bibitem{modellorlc}
 N. Crescini, Understanding photon-magnon hybridization through an analytical
  model (In preparation).

\bibitem{1475-7516-2012-06-013}
P.~Arias, D.~Cadamuro, M.~Goodsell, et~al., Journal of Cosmology and
  Astroparticle Physics \textbf{2012}(06), 013 (2012).
\newblock \urlprefix\url{http://stacks.iop.org/1475-7516/2012/i=06/a=013}

\bibitem{raffeltaxions}
B.B. Kuster~M., Raffelt~G., \emph{Axions. Lecture Notes in Physics} (Springer,
  Berlin, Heidelberg, 2008).
\newblock \urlprefix\url{https://doi.org/10.1007/978-3-540-73518-2_3}

\bibitem{battich}
T.~Battich, A.~Córsico, L.~Gabriel~Althaus, M.~Miguel Miller~Bertolami,
  \textbf{2016}, 062 (2016)

\bibitem{Corsico:2016okh}
A.H. Córsico, A.D. Romero, L.G. Althaus, et~al., JCAP \textbf{1607}(07), 036
  (2016).
\newblock \doi{10.1088/1475-7516/2016/07/036}

\bibitem{PhysRevLett.118.261301}
D.S. Akerib, S.~Alsum, C.~Aquino, et~al., Phys. Rev. Lett. \textbf{118}, 261301
  (2017).
\newblock \doi{10.1103/PhysRevLett.118.261301}.
\newblock
  \urlprefix\url{https://link.aps.org/doi/10.1103/PhysRevLett.118.261301}

\bibitem{201346}
Physics Letters B \textbf{724}(1), 46  (2013).
\newblock \doi{https://doi.org/10.1016/j.physletb.2013.05.060}.
\newblock
  \urlprefix\url{http://www.sciencedirect.com/science/article/pii/S0370269313004425}

\bibitem{PhysRevD.90.062009}
E.~Aprile, F.~Agostini, M.~Alfonsi, et~al., Phys. Rev. D \textbf{90}, 062009
  (2014).
\newblock \doi{10.1103/PhysRevD.90.062009}.
\newblock \urlprefix\url{https://link.aps.org/doi/10.1103/PhysRevD.90.062009}

\bibitem{PhysRevD.95.029904}
E.~Aprile, F.~Agostini, M.~Alfonsi, et~al., Phys. Rev. D \textbf{95}, 029904
  (2017).
\newblock \doi{10.1103/PhysRevD.95.029904}.
\newblock \urlprefix\url{https://link.aps.org/doi/10.1103/PhysRevD.95.029904}

\bibitem{1475-7516-2013-11-067}
E.~Armengaud, Q.~Arnaud, C.~Augier, et~al., Journal of Cosmology and
  Astroparticle Physics \textbf{2013}(11), 067 (2013).
\newblock \urlprefix\url{http://stacks.iop.org/1475-7516/2013/i=11/a=067}

\bibitem{Yoon2016}
Y.S. Yoon, H.K. Park, H.~Bhang, et~al., Journal of High Energy Physics
  \textbf{2016}(6), 11 (2016).
\newblock \doi{10.1007/JHEP06(2016)011}.
\newblock \urlprefix\url{https://doi.org/10.1007/JHEP06(2016)011}

\bibitem{Derbin2012}
A.V. Derbin, I.S. Drachnev, A.S. Kayunov, V.N. Muratova, JETP Letters
  \textbf{95}(7), 339 (2012).
\newblock \doi{10.1134/S002136401207003X}.
\newblock \urlprefix\url{https://doi.org/10.1134/S002136401207003X}

\bibitem{1475-7516-2017-10-010}
M.~Giannotti, I.G. Irastorza, J.~Redondo, A.~Ringwald, K.~Saikawa, Journal of
  Cosmology and Astroparticle Physics \textbf{2017}(10), 010 (2017).
\newblock \urlprefix\url{http://stacks.iop.org/1475-7516/2017/i=10/a=010}

\bibitem{cicoli}
M.~Cicoli, M.D. Goodsell, A.~Ringwald, Journal of High Energy Physics
  \textbf{2012}(10), 146 (2012).
\newblock \doi{10.1007/JHEP10(2012)146}.
\newblock \urlprefix\url{https://doi.org/10.1007/JHEP10(2012)146}

\bibitem{PhysRevD.88.035020}
S.K. Lamoreaux, K.A. van Bibber, K.W. Lehnert, G.~Carosi, Phys. Rev. D
  \textbf{88}, 035020 (2013).
\newblock \doi{10.1103/PhysRevD.88.035020}.
\newblock \urlprefix\url{https://link.aps.org/doi/10.1103/PhysRevD.88.035020}

\bibitem{kuzmin}
L.S. Kuzmin, A.S. Sobolev, C.~Gatti, et~al., IEEE Transactions on Applied
  Superconductivity \textbf{28}(7), 1 (2018).
\newblock \doi{10.1109/TASC.2018.2850019}

\end{thebibliography}
\end{document}